\documentclass[useAMS,usenatbib]{mn2e}
\usepackage{amssymb}
\usepackage{graphicx}
\usepackage{xspace}

\usepackage{amssymb,latexsym,graphicx,natbib,eufrak,times,amsmath}

\usepackage{ifthen}
\def\draftversion{0} % set this to 1 to display the content of the draft commands, or to 0 to hide them.

\makeatletter
\newcommand\mytoc{%
    \@starttoc{toc}%
}
\makeatother

\ifthenelse{\equal{\draftversion}{0}}{
	\usepackage{xcolor}
	\newcommand{\tmp}{}
	\newenvironment{envcomm}[1]{\renewcommand{\tmp}{#1}\begin{color}{blue}\begin{center}\hrule\vspace{0.5mm}\tmp's COMMENTS\end{center}}{\begin{center}END OF \tmp's COMMENTS\vspace{0.5mm}\hrule\end{center}\end{color}}
	\newenvironment{draft}{\begin{color}[rgb]{0,0.4,0}\begin{center}\hrule\vspace{0.5mm}DRAFT\end{center}}{\begin{center}END OF DRAFT\vspace{0.5mm}\hrule\end{center}\end{color}}
	\newcommand{\comcomm}[2]{\begin{color}{blue}\ $\bullet$ \textbf{#1:} #2 $\bullet$\ \end{color}}
	\newcommand{\revend}[1]{\par\begin{color}[rgb]{0,0.4,0}\begin{center}\hrule\vspace{0.5mm}END OF #1's REVISIONS\vspace{0.5mm}\hrule\end{center}\end{color}\par}
	\newcommand{\todo}[1]{\begin{color}{red}\ $\bullet$ \textbf{To do: }#1 $\bullet$\ \end{color}}
	
	\newcommand{\del}[1]{\begin{color}[rgb]{0,0.5,0.0}\ $\bullet$ \textbf{Deleted: }#1 $\bullet$\ \end{color}}
	\newcommand{\sk}[1]{\begin{color}[rgb]{0.6,0,0.6}#1\end{color}}
	\newcommand{\toc}{\par\begin{color}[rgb]{0.6,0,0.6}\begin{center}\hrule\vspace{0.5mm}\begingroup\small\let\cleardoublepage\relax\let\clearpage\relax\mytoc\endgroup\vspace{0.5mm}\hrule\end{center}\end{color}\par}
	}{
	\newsavebox{\trashcan}
	\newenvironment{envcomm}[1]{\begin{lrbox}{\trashcan}\begin{minipage}{\columnwidth}}{\end{minipage}\end{lrbox}}
	
	\newcommand{\comcomm}[2]{}
	\newcommand{\revend}[1]{}
	\newcommand{\todo}[1]{}
	
	\newcommand{\del}[1]{}
	\newcommand{\sk}[1]{}
	\newcommand{\toc}{}
	}
\long\def\symbolfootnote[#1]#2{\begingroup%
\def\thefootnote{\fnsymbol{footnote}}\footnote[#1]{#2}\endgroup} 

\newcommand{\aj}{AJ}% Astronomical Journal
\newcommand{\araa}{ARA\&A}% Annual Review of Astron and Astrophys
\newcommand{\apj}{ApJ}% Astrophysical Journal
\newcommand{\apjl}{ApJ}% Astrophysical Journal, Letters
\newcommand{\aap}{A\&A}% Astronomy and Astrophysics
\newcommand{\aapr}{A\&A~Rev.}% Astronomy and Astrophysics Reviews
\newcommand{\aaps}{A\&AS}% Astronomy and Astrophysics, Supplement
\newcommand{\mnras}{MNRAS}% Monthly Notices of the RAS
\newcommand{\mh}{\ensuremath{\textrm{\,--\,}}}
\newcommand{\bb}[1]{\ifmmode \mbox{\boldmath $ #1$} \else  \mbox{\boldmath $#1$} \fi}

\newcommand{\U}[1]{\ensuremath{\mathrm{~#1}}}
\newcommand{\yr}{\U{yr}}
\newcommand{\Myr}{\U{Myr}}

\newcommand{\pc}{\U{pc}}
\newcommand{\kpc}{\U{kpc}}

\newcommand{\msun}{\U{M}_{\odot}}
\newcommand{\Msun}{\msun}

\newcommand{\cc}{\U{cm^{-3}}}

\newcommand{\kms}{\U{km\ s^{-1}}}

\newcommand{\hii}{H{\sc ii} }

\newcommand{\ramses}{\texttt{RAMSES}\xspace}

\newcommand{\fig}[2][]{Figure#1~\ref{fig:#2}}

\newcommand{\sect}[2][]{Section#1~\ref{sec:#2}}

\renewcommand{\fig}[2][]{Fig#1.~\ref{fig:#2}}

\title[Cloud and star formation in galactic bars]{Environmental regulation of cloud and star formation in galactic bars}

\author[Renaud et al.] {F.~Renaud$^1$\thanks{f.renaud@surrey.ac.uk}, F.~Bournaud$^2$, E.~Emsellem$^{3,4}$, O.~Agertz$^1$,
E.~Athanassoula$^5$, \newauthor F.~Combes$^6$, B.~Elmegreen$^7$, K.~Kraljic$^{5}$, F.~Motte$^2$ and R.~Teyssier$^8$\\
$^1$Department of Physics, University of Surrey, Guildford, GU2 7XH, UK\\
$^2$ Laboratoire AIM Paris-Saclay, CEA/IRFU/SAp-CNRS/INSU, Universit\'e Paris Diderot, F-91191 Gif-sur-Yvette Cedex, France\\
$^3$ European Southern Observatory, 85748 Garching bei Muenchen, Germany\\
$^4$ Universit\'e Lyon 1, Observatoire de Lyon, CRAL et ENS, 9 Av Charles Andr\'e, F-69230 Saint-Genis Laval, France\\
$^5$ Aix Marseille Universit\'e, CNRS, LAM (Laboratoire d'Astrophysique de Marseille), 13388 Marseille, France\\
$^6$ Observatoire de Paris, LERMA, PSL, CNRS, Sorbonne Univ. UPMC and College de France, F-75014, Paris, France\\
$^7$ IBM T. J. Watson Research Center, 1101 Kitchawan Road, Yorktown Heights, New York 10598, USA\\
$^8$ Institute for Theoretical Physics, University of Z\"urich, CH-8057 Z\"urich, Switzerland\\
}

\date{Accepted 2015 September 22.  Received 2015 September 7; in original form 2015 August 10}

\begin{document}
\maketitle

%%%%%%%%%%%%%%%%%%%%%%%%%%%%%%%%%%%%%%%%%%%%%%%%%%%%%%%%%%%%%%%%%%%%%%%%%%%%%%%%

\begin{abstract}
The strong time-dependence of the dynamics of galactic bars yields a complex and rapidly evolving distribution of dense gas and star forming regions. Although bars mainly host regions void of any star formation activity, their extremities can gather the physical conditions for the formation of molecular complexes and mini-starbursts. Using a sub-parsec resolution hydrodynamical simulation of a Milky Way-like galaxy, we probe these conditions to explore how and where bar (hydro-)dynamics favours the formation or destruction of molecular clouds and stars. The interplay between the kpc-scale dynamics (gas flows, shear) and the parsec-scale (turbulence) is key to this problem. We find a strong dichotomy between the leading and trailing sides of the bar, in term of cloud fragmentation and in the age distribution of the young stars. After orbiting along the bar edge, these young structures slow down at the extremities of the bar, where orbital crowding increases the probability of cloud-cloud collision. We find that such events increase the Mach number of the cloud, leading to an enhanced star formation efficiency and finally the formation of massive stellar associations, in a fashion similar to galaxy-galaxy interactions. We highlight the role of bar dynamics in decoupling young stars from the clouds in which they form, and discuss the implications on the injection of feedback into the interstellar medium, in particular in the context of galaxy formation.
\end{abstract}

%\toc

%%%%%%%%%%%%%%%%%%%%%%%%%%%%%%%%%%%%%%%%%%%%%%%%%%%%%%%%%%%%%%%%%%%%%%%%%%%%%%%%
\section{Introduction}

In isolated disc galaxies in the local Universe in general and the Milky Way in particular, most of the star formation activity occurs in giant molecular clouds distributed within kpc-scale structures like spiral arms, where the interstellar medium (ISM) is about one hundred time denser than in inter-arm regions \citep{Visser1980, Hughes2013}. Such excess of star formation is also observed in the central regions of galactic bars, again because of high surface densities of molecular gas (\citealt{Sheth2002}, but see \citealt{Longmore2013} on a deficit of star formation in the central molecular zone). Furthermore, the extremities of bars can host large amounts of molecular gas, distributed in massive complexes, also called giant molecular associations \citep{Solomon1987, Martin1997, Nguyen2011}. Among them, the complex W43 at the tip of the Milky Way bar gathers about $9\times 10^6 \Msun$ of gas, embedded in a neutral hydrogen envelope of $3\times 10^6 \Msun$ \citep{Motte2014}. This complex hosts a mini-starburst activity, with the ongoing formation of massive stars \citep{Motte2003, Louvet2014}. Using position-velocity data, \citet{Motte2014} proposed that such activity results from a collision between clouds accreted from a nearby spiral arm. However, because of the very short dynamical timescales in such environment ($\sim 10 \Myr$) and the interplay between the numerous physical processes at stake, understanding the origin of molecular complexes, their lifetimes, and the physical conditions leading to enhanced star formation in bars is rather difficult observationally. 

Numerical simulations provide a complementary approach to the problem, and have been long used to address the questions related to the dynamics and star formation in bars (see a review in \citealt{Athanassoula2013}). Using sticky particle simulations, \citet{Combes1985} noted asymmetries in the distribution of dust lanes between the leading and trailing sides of bars. Such results have also been established in hydrodynamical models  \citep{Prendergast1983, Athanassoula1992, Piner1995} which also showed that gas concentration on the leading side corresponds to shocks. These simulations have sufficiently high resolutions ($\sim 100 \pc$) to describe the inner structures of the bar. However, such pioneer studies faced technical limitations and could not resolve how these features fragment (or not) into molecular clouds ($\sim 10 \pc$). Although modern simulations still face the long-lasting issue of coupling between the kpc-scale dynamics of the galaxy and the parsec and sub-parsec physics driving star formation and the associated processes like feedback, technical improvements in high performance computation push further the limitations. Present-day galaxy simulations now encompass both scales at once \citep[see e.g.][]{Dobbs2012, Bonnell2013, Fujimoto2014}.

Using the model of the Milky Way presented in \citet{Renaud2013b}, we address here some of the questions related to the evolution of the structure of the ISM and the formation of stars in a galactic bar, focussing on the regions at their tips. The main advantage of this simulation is to describe the evolution of the interstellar medium and the formation of stars by resolving the giant molecular clouds at parsec and sub-parsec scales, within a ``live'' galactic context, i.e. with no forcing of the potential nor of the spiral or bar pattern. Therefore, this simulation encompasses both dynamical aspects from the kpc scales like tides, shear, disc rotation, and local parsec and sub-parsec effects like cloud self-gravity, turbulence and stellar feedback. This paper aims at providing comprehensive scenarios for the formation of gaseous structures, their fragmentation into molecular clouds, the interaction of these clouds, and the potential formation of molecular complexes like W43.

%%%%%%%%%%%%%%%%%%%%%%%%%%%%%%%%%%%%%%%%%%%%%%%%%%%%%%%%%%%%%%%%%%%%%%%%%%%%%%%%
\section{Numerical simulation}

Details of the galaxy simulation used in this work are presented in \citet{Renaud2013b} and summarised here. We have run a hydrodynamical simulation of a Milky Way like galaxy in isolation, using the adaptive mesh refinement code \ramses \citep{Teyssier2002}. The initial conditions are axisymmetric distributions of stars (bulge and discs) and dark matter halo rendered with particles, and gas modelled by an exponential disc on the mesh of the simulation. The stellar component forms spirals and a bar within a few $10^8 \yr$ \citep[see][]{Emsellem2015}. The gas then naturally follows this potential well, and further fragments to form its own structures. The formation and evolution of stellar and gaseous structures naturally arise from the dynamical evolution from axisymmetric initial conditions and are not regulated by a pattern speed nor a fixed potential.

In a cubic box $100 \kpc$ wide, the maximum resolution reached is $0.05 \pc$ for a cloud crossing time (i.e. a few $10^6 \yr$), and sub-parsec resolution for a few cloud lifetimes (i.e. several $10 \Myr$). The equation of state, which is a fit to the actual thermal heating/cooling balance, mimics a hot diffuse halo ($10^{5\mh 7} \U{K}$), an isothermal warm phase ($10^4 \U{K}$ up to $\approx 10^{-1} \cc$) and a polytropic branch at high density, down to $14 \U{K}$. In Milky Way-like galaxies, only a moderate amount of gas lies off the adopted equation, and the heating and cooling processes make it rapidly fall back onto the relation, such that this equation of state provides a good approximation to more elaborate implementations, as shown in \citet{Kraljic2014}. A pressure floor (for gas denser than $\approx 2\times 10^5 \cc$) ensures that the local Jeans length is always resolved by a few cells.

Star formation proceeds at a constant efficiency per free-fall time in the gas denser than $2000 \cc$ (which corresponds, on average over the galaxy, to the transition where self-gravity takes over the supersonic turbulent dominated regime, visible as a power-law tail in the otherwise log-normal density probability distribution function, see \citealt{Renaud2013b}, their Figure 10). The prescription for stellar feedback consists  in various phenomenological recipes for photo-ionisation in \hii regions (by setting the temperature of Str\"omgrem spheres to $2\times 10^4 \U{K}$), radiative pressure (by deposition of radial velocity kicks in these spheres) and kinetic supernova blasts \citep{Renaud2013b}. No magnetic field has been included.

The snapshot at $t = 800 \Myr$ (with $t=0$ corresponding to the initial conditions of the simulation) provides a good match to the observational data of the Milky Way, in term of morphology and kinematics. Although some fine features are absent of the model (e.g. the nuclear ring at $200 \pc$), the main structures of the Milky Way (bar, spiral arms) are well reproduced \citep[see][]{Renaud2013b, Emsellem2015, Chemin2015}. In that sense, although it is still utopian to compare the simulation with observations at the level of individual GMCs, this model can confidently be used to probe the average physical conditions of the real Galaxy and address the questions of formation and evolution of GMCs from a statistical perspective.

\begin{figure}
\includegraphics{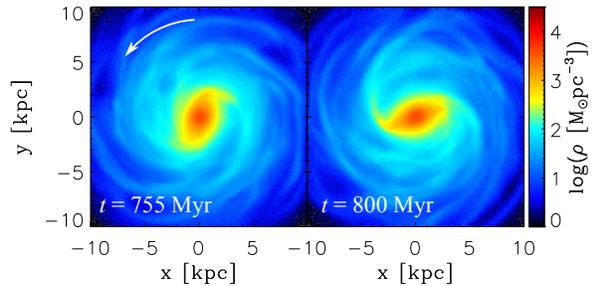}
\caption{Map of the stellar component of our Milky Way model (seen face-on) at the beginning and the end of our period of interest. (The bar rotates anti-clockwise in all our Figures.)}
\label{fig:stellarmap}
\end{figure}

In this contribution, we study the formation of gas clouds and stars in the outer parts of the bar. The (hydro-)dynamics of the innermost kpc, and the coupling between the large-scales and the central black hole are presented in \citet{Emsellem2015}. We focus our analysis between $t = 755 \Myr$ and $t = 800 \Myr$ to monitor the physical conditions leading to the formation of clouds and then to stars (\fig{stellarmap}). Running the simulation at sub-parsec resolution over this period would be too costly. Instead, the resolution is progressively increased, from $3 \pc$ to $0.05 \pc$, one refinement level at a time. Before activating a finer level, we let the simulation run for several free-fall times of the densest structures, to ensure the signatures of previous refinements have been erased. Star formation and feedback are however active throughout the analysed period. Note that the effective resolution controls the fragmentation level of the ISM, which enables us to allow for the formation of spirals and bar while preventing that of molecular clouds during the earliest epochs of the simulation ($\sim 3\mh4$ rotation periods) to avoid a too flocculent disc. The coarsest resolution is however sufficient to describe the properties of the ISM at the scale of molecular clouds (several $10 \pc$), but without being able to describe their inner structures.

The bar formed at $t \approx 600 \Myr$ and evolved from a weak bar (maximum amplitude of the tangential to radial force ratio of $Q \approx 0.2$ at $t = 700 \Myr$) to a stronger bar ($Q \approx 0.6$ at $t = 800 \Myr$, see \citealt{Emsellem2015}, their Section 3.1). At $t=800 \Myr$, the bar pattern-speed is $\approx 58 \U{km/s/kpc}$ (with co-rotation found at $3.6 \kpc$). Its semi-major axis is $2.4 \kpc$, with an axis ratio of 1:3. Inner Lindblad resonances, as computed from the azimuthally averaged mass distribution, are found at $40 \pc$ and $450 \pc$ from the galactic centre. More details on the evolution of the disc and the bar can be found in \citet{Renaud2013b} and \citet{Emsellem2015}.

%%%%%%%%%%%%%%%%%%%%%%%%%%%%%%%%%%%%%%%%%%%%%%%%%%%%%%%%%%%%%%%%%%%%%%%%%%%%%%%%
\section{Results}

\begin{figure}
\includegraphics{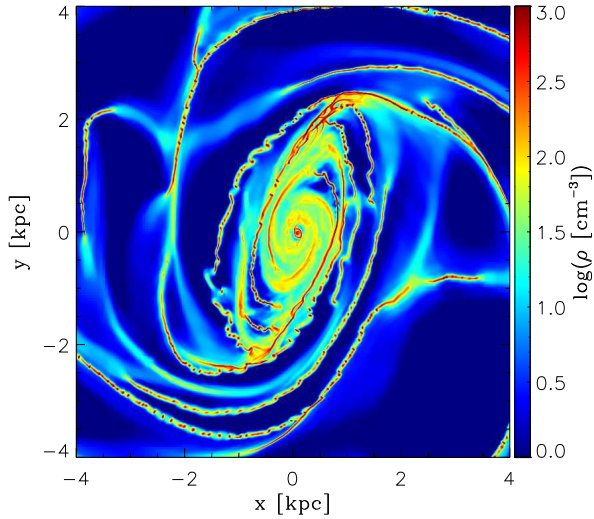}
\caption{Map of the gas density (at $t = 755 \Myr$), in the bar and the inner region of the disc.}
\label{fig:density}
\end{figure}

\fig{density} shows the density of gas at $t = 755 \Myr$, before most of the star forming clumps form in the bar. Using this snapshot allows us to measure the physical conditions of cloud formation. Gas has already gathered in dense structures, following the gravitational potential of the stellar component, and is at the turning point of fragmenting into clouds. Some regions along the outermost spiral arms (e.g. on lower part of \fig{density}) already host beads-on-a-string clouds \citep{Elmegreen1983, Renaud2013b}, which are still in the phase of accreting their surrounding material. Gravitational torques from the bar and resonances regulate the gas flows toward the centre along mini-spirals, in particular inside the inner kpc, as discussed in \citet{Emsellem2015}.

%%%%%%%%%%%%%%%%%
\subsection{Turbulence}
\label{sec:turbulence}

\begin{figure}
\includegraphics{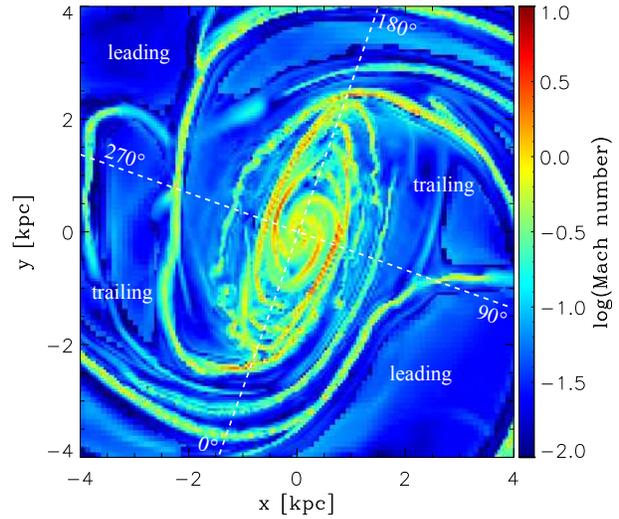}
\caption{Map of the turbulent Mach number evaluated at the scale $24 \pc$. Dashed lines indicate the major and minor axes of the bar, and thus tell apart the leading and trailing quadrants. The highest turbulent Mach numbers are found on leading edges of the bar, where the ISM gets supersonically shocked.}
\label{fig:mach}
\end{figure}

\begin{figure}
\includegraphics{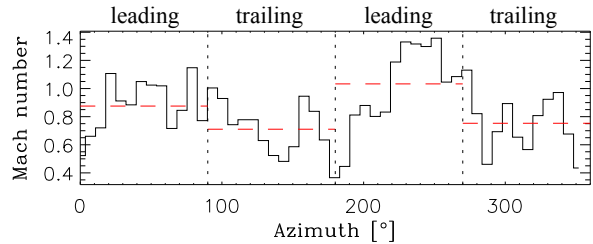}
\caption{Azimuthal profile of the turbulent Mach number (computed at the scale of $24 \pc$), along the edge of the bar. The edge is here defined as an ellipse of semi-major and semi-minor axes $1.8 \kpc$ and $0.7 \kpc$ respectively, and with an inclination of 72 degrees with respect to the axis $x=0$ (see \fig{mach}). To account for the non-perfectly ellipsoidal nature of the bar and include the densest gas features, data is extracted in a ring spanning $250 \pc$ on both the inner and outer sides of this ellipse. Azimuths are indicated on \fig{mach}. On each quadrant (matching those displayed in \fig{mach}), a horizontal dashed line indicates the average turbulent Mach number. The leading sides of the bar host regions with higher turbulent Mach numbers than the trailing quadrants.}
\label{fig:mach_profile}
\end{figure}

At the scale of $24 \pc$, only a few regions yield supersonic turbulence (\fig{mach}): the leading edges of the bar (as observed by \citealt{Sheth2002}) and, to a lower extent, the two main spiral arms close to the points where they connect to the bar. (The turbulent Mach number is computed as in \citealt{Kraljic2014}, their Section 4.1.) To further illustrate this point, \fig{mach_profile} displays the azimuthal profile of the turbulent Mach number along the edge of the bar, confirming the presence of a leading versus trailing asymmetry.

These high turbulent Mach numbers mostly originate from a velocity dispersion ($\sim 20 \kms$) about 2 to 10 times larger in these regions than elsewhere along the bar edge (where density, temperature, and thus sound speed are comparable). The locations of these regions suggest the presence of a supersonic shock between the edge of the rotating bar ($\sim 60 \U{km/s/kpc}$, $50 \mh 500 \U{K}$, $\sim 10^{2\mh 4} \cc$) and the warm neutral medium ($2500 \mh 3000 \U{K}$, $\sim 10 \cc$) surrounding it. (Note that hotter and more diffuse inter-arm medium is found at larger radii, and only faces the spirals and not the bar itself.) Oppositely, the inner region of the bar stands in transonic and subsonic regimes.

Number of studies revealed that, on top of the turbulent Mach number which describes the level of turbulence of the ISM, the nature of the turbulence can also be key in the formation and evolution of dense gas structures (see e.g. \citealt{Wada2002, Kritsuk2007, Federrath2010} at cloud-scales, and \citealt{Renaud2014b} at galactic scales). This complementary aspect can be probed by measuring the relative importance of the compressive (curl-free) and solenoidal (divergence-free) modes of turbulence. To do so, we follow the method presented in \citet[their Appendix A]{Renaud2014b}. The divergence and curl of the velocity field are computed with first-order finite differences using a group of $2^3$ grid cells. In such scheme, the large scale component of the total velocity field (i.e. the non-turbulent term) cancels out, and only the turbulent components of the velocity field remain. By multiplying the divergence (respectively curl) by the scale-length over which it is evaluated (i.e. the cell size), one reconstructs the compressive (respectively solenoidal) velocity, which then can be expressed as a kinetic energy. Note that the curl-free component represents both a compression and a rarefaction effect, but we refer to it as compressive mode, for short.

\begin{figure}
\includegraphics{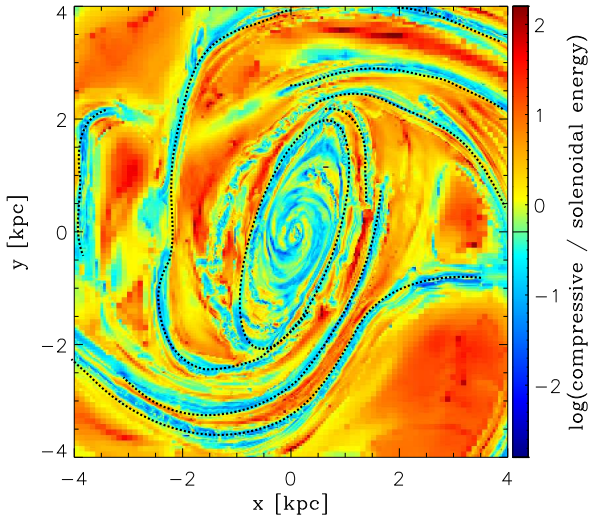}
\caption{Ratio of the compressive turbulence energy (encompassing both compression and rarefaction) to the solenoidal turbulence energy, both evaluated at the scale of $24 \pc$. The dotted ellipse and lines indicate the approximate location of the densest gas structures along the edge of the bar and the spiral arms. The horizontal mark in the color bar indicates energy equipartition between the two turbulent modes (i.e. a ratio of 1/2, see text for details).}
\label{fig:solenoidal}
\end{figure}

\fig{solenoidal} maps the ratio of the energies of the two turbulent modes. Because the compressive mode carries one degree of freedom while the solenoidal mode has two, energy equipartition between the two modes is reached when their ratio equals one half. The discrete nature of the grid on which the motions are represented implies that this ratio is over-estimated if evaluated over too few resolution elements \citep[see][]{Federrath2011}. By varying the resolution at which we compute the map of \fig{solenoidal}, we ensured that convergence is reached at $24 \pc$, i.e. over at least 8 cells\footnote{In the lowest density regions (e.g. the inter-arm), the adaptive nature of the grid refinement implies that the cell size is larger than in denser zones like the bar and the spiral arms. Because of this, at fixed scale, the turbulent modes are computed over fewer resolution elements, and thus the importance of the compressive mode there (as a rarefaction effect) is over-estimated. In our case, only the extreme values in over-compressive regions (of very low density) are significantly changed (up to a factor of 3 in the most extreme cases), but the overall behaviour remains unaffected if quantities are computed at $100 \pc$ scale instead of $24 \pc$.}.

Sharp density gradients, across the spiral arms and at the edge of the bar, are often associated with a change from compressive- to solenoidal-dominated turbulence. Most of the dense structures (bar and spirals) yield turbulence close to equipartition, or slightly solenoidal dominated, indicating the presence of local shear motions. Only the low density gas (mostly in the inter-arm regions) yield a compressive-dominated turbulence (rarefaction) over kpc-scale volumes. Recall that at this stage of the simulation, star formation has not taken place yet and thus, this turbulence is not generated by feedback, but only by the large-scale (hydro-)dynamics. Later, the turbulence pumped by stellar feedback at parsec scales will combine with that injected at kpc-scale by the galactic dynamics.

Along the bar edge, the leading versus trailing dichotomy noted in \fig{mach} and \ref{fig:mach_profile} for the Mach number is not recovered in the nature of turbulence. The regions along the edge share comparable turbulence as the spiral arms. No major differences are found between the edge of the bar and the central kpc. The main differences in the organisation of the ISM is thus originating from large-scale aspects like shear (see next Section).

Therefore, at the scale of molecular clouds, the overall turbulence intensity (Mach number) shows a rather strong dependence on the galactic environment, leading to differences in the structure of the ISM and the further formation of dense clouds. The nature of turbulence (compressive versus solenoidal) in the dense gas structures is however fairly independent of the kpc-scale structures, contrarily to interacting galaxies where it is mostly compressive \citep{Renaud2014b}. The reason for this might be that compressive tides, which are thought to drive large-scale compressive turbulence in interacting galaxies, do not exist over kpc-scale volumes in isolated galaxies.

This picture remains valid at later stages in the simulation, even after the formation and collapse of clouds.

%%%%%%%%%%%%%%%%%
\subsection{Large-scale velocity field}
\label{sec:velocity}

Orbital motions inside the bar induce a shearing effect on the ISM \citep[see][]{Athanassoula1992, Emsellem2015}. Such shear prevents the formation of dense structures like clouds, which later translates in a lack of star formation \emph{inside} the bar, despite relatively high densities. At the edge of the bar however, supersonic turbulence (as discussed above) balances a weaker shear (and tides) than in the innermost regions, and favours the formation and survival of overdensities.

\begin{figure}
\includegraphics{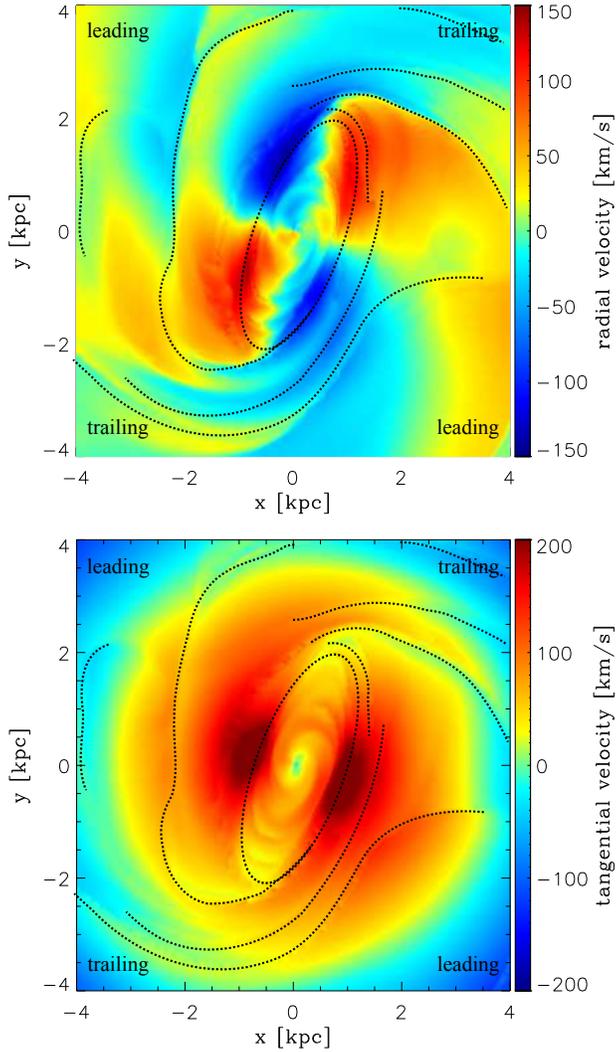}
\caption{Top: gas radial velocity. Blue and red indicate inward and outward motions respectively. Bottom: gas tangential velocity, minus the pattern speed of the bar ($58 \U{km/s/kpc}$). Blue and red indicate clockwise and anti-clockwise motions respectively. The dotted ellipse and lines indicate the approximate location of the densest gas structures along the edge of the bar and the spiral arms.}
\label{fig:velocity}
\end{figure}

\fig{velocity} shows the maps of the radial and tangential components of the gas velocity in the reference frame of the bar. The highest radial velocities are found along the edge of the bar (pointing inward on the leading sides and outward on the trailing sides). Shear is clearly visible perpendicular to the bar major and minor axes (in particular in radial velocity) whereas kpc-size regions along the edge show almost uniform velocities. In (relative) absence of large-scale shear in the latter areas, small-scale ($< 100 \pc$) motions dominate the velocity field, which translates into the turbulence patterns discussed in the previous Section.

\begin{figure}
\includegraphics{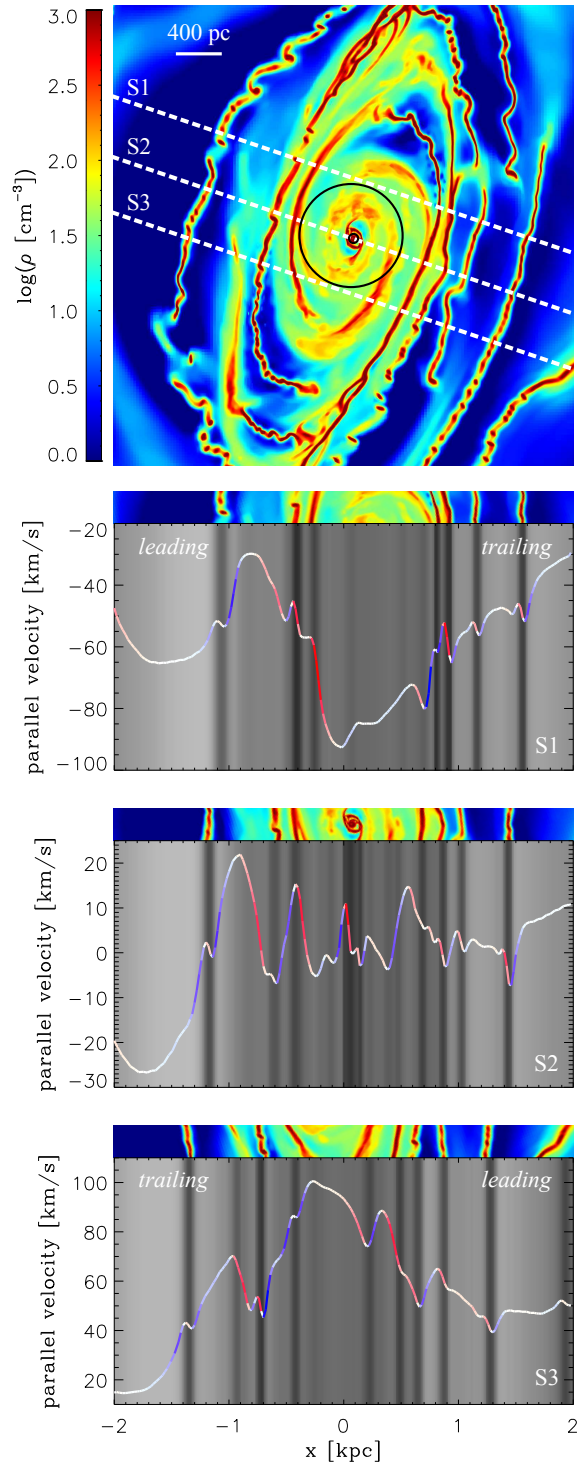}
\caption{Top: gas density map of the bar (as in \fig{density}). The black circles show the inner Lindblad resonances ($40 \pc$ and $450 \pc$). The three dashed lines indicate the position of ``slits'' parallel to the minor axis, and shifted along the vertical axis by $500 \pc$, $0 \pc$ and $-500 \pc$. The bottom three panels show the velocity component parallel to the minor axis, along the three slits (with the convention of positive velocity representing motion from left to right). To ease the interpretation, the background greyscale image represents the logarithm of the gas density and the curve colour codes the local velocity gradient. With the sign convention adopted, red indicates a converging flow, blue a diverging flow, white and symmetric features denotes bulk motions. A density map of the corresponding region is shown above each panel to help the identification of structures.}
\label{fig:velocity_profile}
\end{figure}

\fig{velocity_profile} displays the velocity component parallel to the bar minor axis along three slits, following the approach of \citet[her Figure 11]{Athanassoula1992}. In the outermost regions considered here, over-densities correspond to spiral arms. Any of the associated velocity profiles \emph{within} the arm (i.e. over $\sim 100 \pc$) is monotonically decreasing, which indicates converging flow toward the arm centre. At larger scale, the flows are set by the disc dynamics and thus vary with location.

The innermost dense gas structures (only visible in the central slit) match the positions of the inner Lindblad resonances, at $40 \pc$ and $450 \pc$. The gas is fuelled there along mini-spiral arms by the gravitational torques of the bar and piled up on these stable orbits \citep[see][]{Renaud2013b, Emsellem2015}. Regions in between these orbits yield either low velocities or diverging flows, which participate in the lack of cloud and star formation in this area.

\begin{figure}
\includegraphics{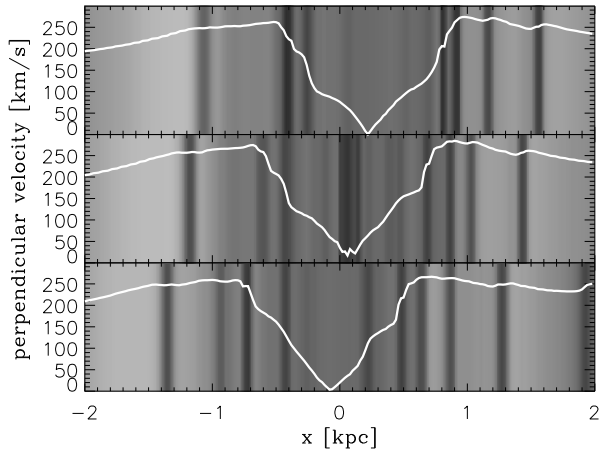}
\caption{Profile of the velocity component perpendicular to the minor axis, along the three slits shown in \fig{velocity_profile}. Background indicates the gas density in the slits, as in \fig{velocity_profile}.}
\label{fig:shear_profile}
\end{figure}

At the edge of the bar ($0.5 \kpc \lesssim |x| \lesssim 1 \kpc$), the densest structures match steep velocity gradients, i.e. shocks, especially on the leading side of the bar ($x \approx 300 \pm 100 \pc$ for S1 and $x \approx 600 \pm 100 \pc$ for S3), which confirms the findings of \citet{Athanassoula1992}. On the trailing side, the velocity profile is mostly symmetric over a few $100 \pc$, which simply denotes the rotation of the bar.

However, structures at intermediate radii and on the trailing side are not systematically associated with shocks. The difference originates from the velocity component perpendicular to the minor axis of the bar, of which profile is displayed on \fig{shear_profile}. Dense gas structures associated with shocks (i.e. with steep parallel velocity profiles on \fig{velocity_profile}) are found in zones of low shear (i.e. flat perpendicular velocity profiles \fig{shear_profile}), particularly close to the edge of the bar, and in a few $\sim 100 \pc$-scale regions where the perpendicular velocity profile locally flattens.

Such relatively weak shear, combined with high densities and strong turbulence (recall \sect{turbulence}), shapes elongated dense gaseous structures and allows for their survival until they fragment into clouds. Most of the material found there does not spiral inside the bar but remains on the bar edge and circulates around the bar (on $x_1$-type orbits, see \citealt{Contopoulos1989, Skokos2002}) at high velocity (up to $230 \kms$ in the bar reference frame). It takes about $20 \Myr$ to travel from one of its apocentre to the next (on the other side of the bar, $\approx 3 \kpc$ on average). The eccentric nature of this motion makes the gas slow down ($\sim 40 \kms$) when it reaches the apocentre, i.e. the extremities of the bar (along the major axis). Here, the spread in orbital energies translates into a radial spread of apocentres, such that the material coming from the thin filaments parallel to the bar spans a large area (about $1 \kpc^2$, between $\sim 1 \kpc$ and $\sim 2 \kpc$ away from the galactic centre). Despite this large volume in space, the slowing down of gas clouds makes them accumulate there (before being accelerated again along the oposite bar edge). We identified about 50 dense clouds ($> 2000 \cc$) at each extremity of the bar, with a velocity dispersion between them of $\sim 40 \kms$. This accumulation in space and time, so-called orbital crowding \citep{Kenney1991}, increases the probability of cloud-cloud collisions, as already noted by \citet{Rodriguez2008}. Such events lead to star formation, as discussed in \sect{ccc}.

%%%%%%%%%%%%%%%%%
\subsection{Young stars}

The natural consequence of the formation of dense clouds is that self-gravity will eventually take over regulating mechanisms and will lead to cloud collapse and star formation. 

\begin{figure}
\includegraphics{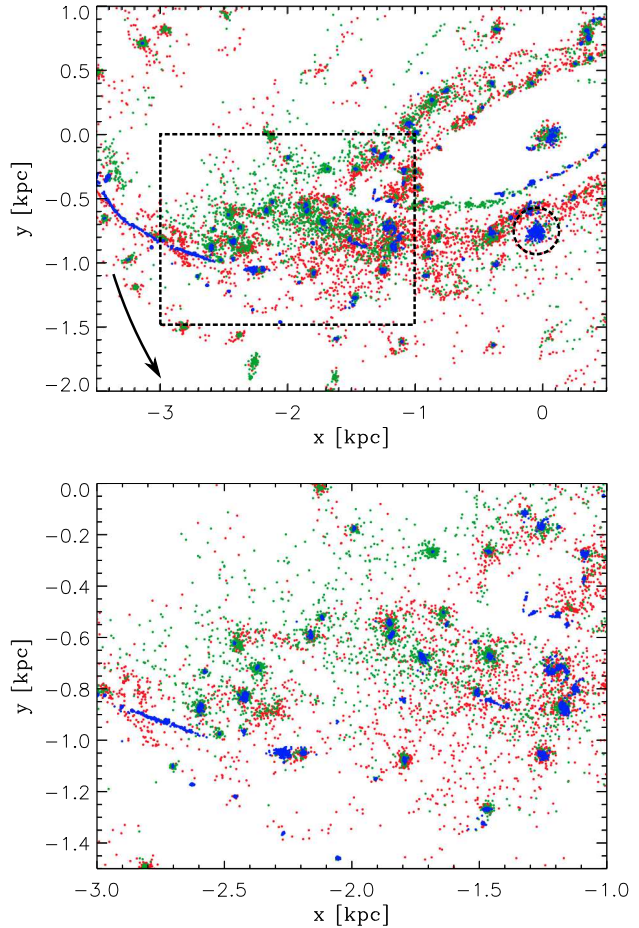}
\caption{Age of the young stars in the vicinity of the tip of the bar at $t = 800 \Myr$ (i.e. $\approx 45 \Myr$ after the instant showed in \fig{density}). Only one side of the bar is shown, for the sake of clarity. An arrow indicates the rotation of the bar. The object circled is discussed in \sect{ccc}. The bottom panel shows a zoom-in of the rectangle area in the top panel. We consider three age bins: $< 20 \Myr$ (blue), between $20$ and $40 \Myr$ (green), and $> 40 \Myr$ (red).}
\label{fig:age}
\end{figure}

\fig{age} shows the spatial distribution and the age of the new stellar particles (i.e. the stars formed out ot gas during the simulation), about $45 \Myr$ after the instant previously discussed, once the clouds have formed, collapsed and began to form their stellar material. The lack of star formation in the inner bar (except at the very centre) is caused by the strong shear destroying clouds in this region (\citealt{Emsellem2015}, see also \fig{shear_profile}), and the orbital pattern preventing stars formed elsewhere to enter this area. 

One can tell apart three stellar populations: (i) the very young stars ($< 20 \Myr$), (ii) the intermediate class (between $20$ and $40 \Myr$ old), and (iii) the older population ($> 40 \Myr$). (These classes are defined arbitrarily, only for the sake of clarity.)

As expected, the very young stars are found associated with dense gas structures: along the spiral arm connected to the tip of the bar (i.e. in the elongated blue feature on the left-hand side of the figure), and in central regions of some dense clumps at the tip of the bar itself (bottom panel), with the notable exception of the massive clump on the edge of the bar ($x \approx -0.1 \kpc$, $y \approx -0.8 \kpc$) which will be discussed in \sect{ccc}.

\begin{figure}
\includegraphics{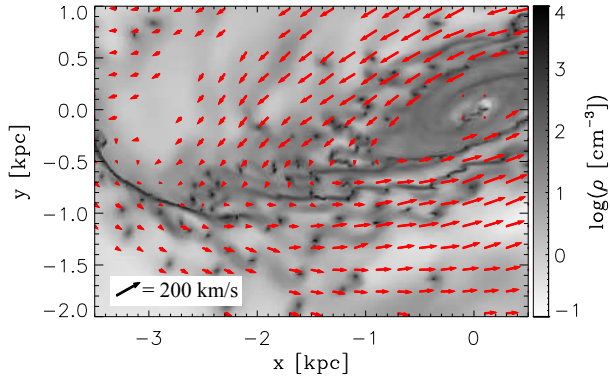}
\caption{Velocity field of the young stars, in the reference frame of the bar ($58 \U{km/s/kpc}$), in the same region and at the same epoch as in \fig{age}. The background displays the gas density map. Like the clouds, the stars slow down when reaching the extremity of the bar.}
\label{fig:velocity_quiver}
\end{figure}

The intermediate age stars ($20 \Myr <$ age $< 40 \Myr$) are preferentially found at the tip of the bar, but on the trailing side (upper half of the bottom panel of \fig{age}). Their ages indicate that they formed at the opposite tip of the bar and have travelled along the (upper) edge of the bar within the last $\sim 20\mh 40 \Myr$, at the high velocity noted above. When reaching this extremity of the bar (i.e. their orbital apocentre), these stars slow down, as shown in \fig{velocity_quiver}. They stay in the area of the tip of the bar for $\sim 20 \Myr$ before being slingshot along the leading side of the bar (bottom on \fig{age} and \ref{fig:velocity_quiver}) toward the other tip. This explains that stars on the leading side of the tip are a couple of $10 \Myr$ older than those on the trailing side: they aged in the area of the tip. 

By tracking back the origins of the stars found at the extremity of the bar (by selecting all stars in the bottom panel of \fig{age}), we find that about 50\% of them form in spiral arms and are accreted later by the bar, with the furthest formation site being $4.5 \kpc$ away from the galactic centre. Among this accreted population, about half ($55 \%$) originates from the main spiral arms connected to the bar, while the other half forms along a secondary structure, equivalent to the 3-kpc arm (or Norma arm) in the real Galaxy. While these stars form in clusters, the dynamics of the accretion (shear and tides) makes most of these objects dissolve before they reach the extremity of the bar. They participate in populating the ``field'', as opposed to the clustered distribution of the young stars formed \emph{in situ}, as visible on \fig{age}.

To summarise, the highly dynamical environment of the galactic bar directly influences the spatial distribution of young stars. It is however likely that the stellar populations will mix over several orbital periods ($\sim 100 \Myr$) and that the long-term evolution of the bar and the associated orbits would amplify this mixing process.

%%%%%%%%%%%%%%%%%
\subsection{Cloud-cloud collisions and W43-analogue}
\label{sec:ccc}

\begin{figure}
\includegraphics{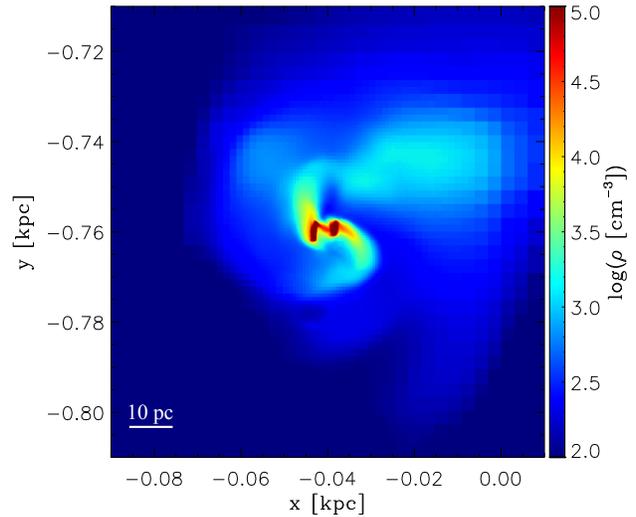}
\caption{Gas density map in the region of interacting clouds along the edge of the bar (at $t = 800 \Myr$). Signatures of cloud-cloud interaction are visible, as tidal tails and bridges made of dense gas.}
\label{fig:ccc}
\end{figure}

The cluster of young stars at the position $x \approx -0.1 \kpc$, $y \approx -0.8 \kpc$ in \fig{age} stands out, both in terms of age ($5 \Myr$ on average) and mass ($3\times 10^6 \Msun$). The on-going star formation in this region contrasts with the surrounding stellar material that has formed either at the tip of the bar a few $\sim 30\mh 40 \Myr$ before, or even earlier elsewhere along the bar or in the spiral arms. The reason for this is the interaction of two gas clouds in this region. \fig{ccc} shows the distribution of gas, in the densest regions of the clouds, which exhibits tail features. These tails are a signature of the past and still on-going tidal interaction between the clouds, in a very similar fashion as interacting galaxies exhibit long tails at much larger scale \citep[e.g.][]{Duc2013}. 

\begin{figure*}
\includegraphics{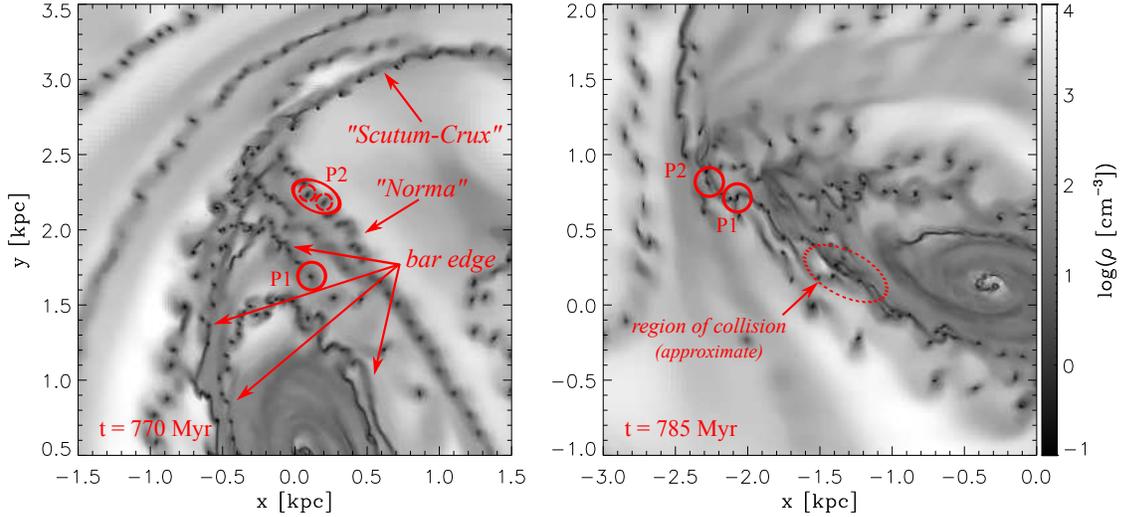}
\caption{Positions of the cloud progenitors of the cloud-cloud collision, $30 \Myr$ (left) and $15 \Myr$ (right) before the instant showed in \fig{ccc}. The ``P1'' object is formed along the edge of the bar, while the other one (``P2'') originates from a beads-on-a-string structure in one spiral arm. The two meet at the extremity of the bar. The progenitor ``P2'' is itself the result of the collision of smaller clouds that occurs between the two moments pictured here. The Scutum-Crux and Norma spiral arms are visually identified with those of the real Galaxy, only for the sake of clarity. The label ``bar edge'' points to elongated gas structures associated to the bar, i.e. not spirals.}
\label{fig:tracking}
\end{figure*}

The Eulerian nature of the code used makes it difficult to track the cloud progenitors back. However, by tracking back the stellar particles found in this structure, we are able to get insights on the formation of this object. \fig{tracking} shows the identification of the clouds in earlier snapshots, respectively 30 and 15 Myr before the instant of \fig{age} and \fig{ccc}. The two progenitors visible in \fig{ccc} are labelled ``P1'' and ``P2'' here. (Distinguishing which is which on \fig{ccc} is impossible because of a too low time sampling of the simulation snapshots compared to their orbital timescale around each other, and because their stellar contents are too spatially extended, such that they overlap on the late snapshot.)

P2 is itself the result of a merger of (at least two, but possibly more) beads-on-a-string clouds that formed along a spiral arm. By visual identification with the real Galaxy, these clouds would be found in the Norma arm, as labelled on \fig{tracking}. Approximately when the P2 structure crosses the bar major axis, at its extremity, orbital crowding makes its two components interact and merge. A few percent of the stellar content of the final object (\fig{age}) is formed during this event, but a significant fraction of the stellar production gets unbound and gets mixed with field stars. This is due to the time-varying potential, tides, asymmetric drift to a lower extent, and the internal evolution of the stellar association (although the collisionless aspect of the code and the softening of the gravitational potential at the resolution of the grid introduce biases in the binding of the stellar structures).

A few Myr later, the resulting P2 is found on an orbit very close to that of P1, another cloud that formed along the edge of the bar. The two progenitors are then accelerated along the leading side of the bar (bottom on the right-panel of \fig{tracking}) with a relative velocity of $85 \kms$. While leaving the tip of the bar, the clouds gravitationally capture each other and interact for the first time ($t = 795 \Myr$), $5 \Myr$ before the instant of \fig{ccc}, which corresponds to the mean age of the stars in the resulting cluster (\fig{age}). The combined mass of the two clouds reaches $6 \times 10^6 \Msun$, i.e. the regime of molecular complexes. The relative velocity of the two progenitors is higher than that observed in the Milky Way spiral arms \citep{Fukui2014}, highlighting again the peculiar dynamics of the tips of the bar.

\begin{figure}
\includegraphics{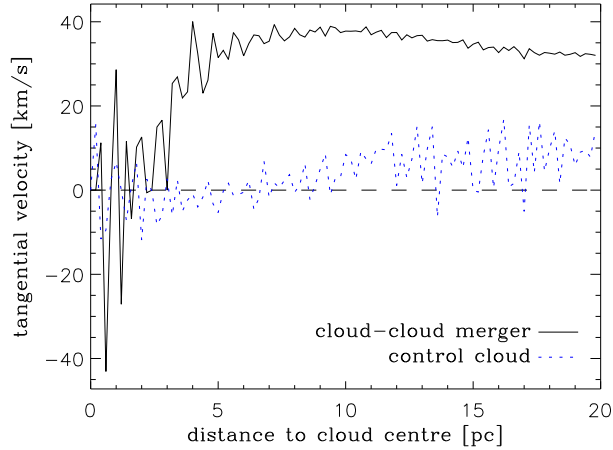}
\caption{Tangential velocity of the gas surrounding the cloud-cloud merger shown in \fig{ccc}, compared to that of a ``control'' cloud which has not experienced any recent interaction. Distances and velocities are computed in the reference frame of the cloud centre of mass. The merger exhibits a rapid rotation (with a complex behaviour in the central $\sim 3 \pc$), induced by the cloud-cloud interaction, while the rotation of the control cloud is several times slower.}
\label{fig:cloud_rotation}
\end{figure}

\fig{cloud_rotation} shows that the interaction of the two progenitors induces a rapid rotation of the merger (anti-clockwise on \fig{ccc}) in their orbital plane, i.e. the plane of the galactic disc. The innermost $\sim 3 \pc$ of the merger host a complex velocity field resulting from the overlap of the internal fields of the two dense structures visible in \fig{ccc}, as well as that of the tidal bridge connecting them. At larger distances, we find tangential velocities of the order of $30 \kms$ (and locally up to $100 \kms$), followed by a Keplerian decrease up to the point ($\sim 100 \pc$) where kpc-scale dynamics takes over in setting the velocity field. Conversely, a ``control'' cloud of comparable central density, in the same environment, and which has not experienced any recent interaction, does not exhibit such rapid rotation. Instead, motions around the control cloud are mainly radial ($\sim 20 \kms$), reflecting the on-going accretion of surrounding gas.

During the cloud-cloud collision event, the increase in density and pressure triggers the bulk of star formation over $\sim 20 \pc$, while the densest parts of the clouds continue spiralling around each other (as visible on \fig{ccc}). This agrees with the observational and theoretical works which showed that such events could trigger an episode of star formation \citep[among others, see][]{Loren1976, Tan2000, Furukawa2009, Tasker2009, Inoue2013}. On top of increasing the star formation rate, collisions are also suspected to gather the physical conditions for the formation of massive stars \citep{Anathpindika2010, Takahira2014, Motte2014}. Although this aspect cannot be probed in our simulation, it suggests that the kpc-scale dynamics of the bar, through cloud-cloud collisions, could lead to the formation of massive stars in excess with respect to those formed in other, more quiescent environments. This excess could either reflect a size-of-sample effect in which the production of more stars overall is likely to sample out further into the high-mass tail of the initial mass function (IMF), or it could reflect a real change in the IMF toward a statistically significant increase in the ratio of high to low mass stars.

\begin{figure}
\includegraphics{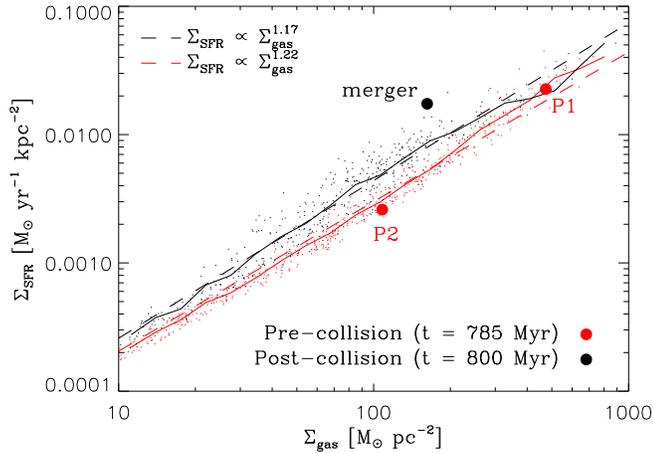}
\caption{Schmidt-Kennicutt diagram of all the clouds in the simulation, highlighting the progenitors and their merger of the cloud-cloud collision discussed in the text. Surface densities are computed using the dense phase only ($> 2000 \cc$) at the scale of $50 \pc$. The dashes lines indicate power-law fits to all the clouds, and the solid lines show the average values.}
\label{fig:ks}
\end{figure}

\fig{ks} shows a Schmidt-Kennicutt diagram of the pre- and post-collision stages, comparing the progenitors and the merger to all the clouds in the simulation. By applying a density selection\footnote{Note that we use this approach to minimize the numerical noise and the effect of star formation stochasticity, which become significant at the scale of individual clouds in our simulation. By using a density threshold and thus focussing on the densest regions of molecular clouds ($\sim 5 \mh 10 \pc$), our method differs from observational methods where a beam of fixed size is used, which explains slight differences in the normalisation of the relations shown in \fig{ks} and the literature. See \citet{Kraljic2014} for a more observation-oriented approach.} above $2000 \cc$, we find that, before the collision, P1 is among the most massive clouds ($2 \times 10^6 \Msun$), while P2 is closer to average ($4\times 10^5 \Msun$). Their Mach numbers are comparable with that of the bulk of the clouds, i.e. in the transonic regime ($\approx 0.8$ at the scale of $24 \pc$). After the P1-P2 collision, most of the clouds in the simulation have fragmented further and formed their star clusters. This translates into a mild shift of the Schmidt-Kennicutt relation toward higher surface densities of gas and SFR. The remnant of the P1-P2 collision weights $5\times 10^6 \Msun$ (in dense gas), and yields a higher Mach number than its progenitors ($\approx 1.3$). This translates in the Schmidt-Kennicutt diagram as the merger hosting a more efficient star formation (i.e. a shorter gas depletion time) than average. This confirms the predictions of \citet{Renaud2012} that supersonic turbulence increases the efficiency of the conversion of gas into stars or in other words, decreases the gas depletion time \citep[see also][]{Kraljic2014}. We note that such evolution is comparable to that of galaxies jumping from the regime of isolated discs to that of starburst during a galaxy-galaxy interaction (\citealt{Renaud2014b}, see also \citealt{Daddi2010b})\footnote{Pushing the analogy further would require to check that the turbulence of the merger is dominated by its compressive mode. However, the resolution of our simulation forbids such analysis at the scale of molecular clouds.}. We also note that the morphology of collapsing beads-on-a-string seen in the spiral arms in our simulation suggests that most star formation over the galactic disc is spontaneous and not triggered by a collision between two large clouds.

In term of complex formation history, extreme mass, and because of its location close to the tip of the bar, our object resembles the observed complex W43 which is thought to originate from multiple cloud accretion events. In particular, \citet{Motte2014} proposed that W43 could have been formed through the accretion of clouds form the nearby Scutum-Centaurus spiral arm. \citet{Carlhoff2013} noted a strong rotation of the main cloud of the W43 complex, together with elongated, arm-like, structures. These are further hints of a recent collision, in line with the formation scenario proposed from our simulation. We note that the orbit of our W43-analogue maintains it along the edge of the bar and at the tips, so that it does not experience the strong shear seen inside the bar. This confirms the suggestion from \citet{Carlhoff2013} that W43 is stable against shear, and could survive several $10 \Myr$ before being destroyed by stellar feedback or tidal effects.

Despite a few differences\footnote{In particular, the observed object is younger ($1 \Myr$) than the simulated one, and found closer to the extremity of the bar.}, the formation of our object yields remarkable similarities to that of W43. In that sense, without providing a perfect match, our simulation would picture a W43-analogue, confirming the possibility of the scenario invoking accretion and collision of (at least) two clouds. In our simulation, some events taking place before the main collision itself have led to the formation and ejection of several loosely bound or unbound stellar associations over the last $\sim 30 \Myr$. A fraction of this stellar population might be identified as the red super giant clusters (RSGC) detected in the vicinity of W43 \citep{Negueruela2010, Negueruela2012, Gonzalez2012a, Gonzalez2012b}. This would then confirm the hypothesis of \citet{Motte2014} that these observed objects ($\approx 20 \Myr$-old) originate from previous star forming events in W43 and have decoupled from the molecular complex at an early stage. A more detailed analysis of this point would require a collisional treatment of gravitation, which is not included in our simulation.

Although cloud-cloud interaction are frequent at the tip of the bar due to orbital crowding, the formation of an object as massive as W43 remains a rare event. In particular, we have not found a comparable structure at the other extremity of the bar despite a strong symmetry of the large scale galactic features (bar and spiral arms). The same situation occurs in observations, where no W43-equivalent has been detected at the remote extremity of the bar.

%%%%%%%%%%%%%%%%%
\subsection{Cloud-star decoupling}

%%%%%
\subsubsection{Process}

\begin{figure}
\includegraphics{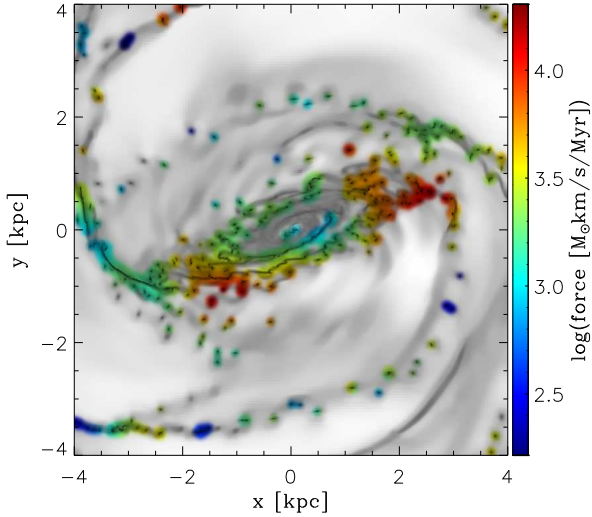}
\caption{Map of the decoupling force between the stars and the gas clouds, overlaid on the gas density map (in greyscale) at $t=800 \Myr$. Red areas indicate that stars are efficiently decoupled from the cloud in which they formed. (See text and Appendix~\ref{app:decoupling}.)}
\label{fig:decoupling}
\end{figure}

\citet{Bash1977} and \citet{Bash1979} found that stars remain associated to the cloud in which they formed for $30\mh 40 \Myr$. After this, they decouple from the gas structures, either because the cloud gets destroyed by feedback, or because large-scale dynamics makes the orbits of the stellar and gas component diverge. Such effect is indeed present in our simulation, as an asymmetric drift of young stars in the spiral arms (see \citealt{Renaud2013b}, their Section 4.5). In the vicinity of the bar, i.e. at smaller galactic radii, the time-scale of the asymmetric drift is shorter, and other dynamical aspects like shear and tides must also be taken into account in the decoupling process.

To understand to what extent young stars can be used to probe the physical conditions of their formation sites, we measure the intensity of a ``decoupling force'' that we define as the velocity of a young star ($< 10 \Myr$) with respect to its closest cloud, divided by the age of the star and multiplied by its mass (see Appendix~\ref{app:decoupling} for details). \fig{decoupling} shows that the decoupling force is the strongest at the tips of the bar and, to a lower extent, stronger along the edge of the bar than in spiral arms, as expected. 

On top of the usual decoupling effect due to the dissipative nature of the gas, kpc-scale dynamics accelerates the drift of stars away from their birth places. Along the edge of the bar (parallel to the major axis), the small galactocentric distance implies an enhanced effect of tides. However, the higher density of (massive) clouds and star clusters due to orbital crowding at the extremities of the bar also induces a significant cloud-cloud tidal effect (with signatures as those shown in \fig{ccc} in the most extreme cases). The tidal force\footnote{computed as first order finite differences of the gravitational force at the scale of $24 \pc$.} is locally up to six times stronger in between the nearby clouds at the tips of the bar than along the bar edge. 

Shear in the dense regions is of comparable amplitude at the extremities and along the edge of the bar, but varies significantly when considering less dense media ($\lesssim 500 \cc$), as noted by \citet{Emsellem2015}.

Finally, the combination of locally strong tides, shear and the dissipative nature of the gas enhances significantly the decoupling of stars from their cloud progenitors at the tips of the bar, with respect to any other region in the galaxy. Young stars are thus likely to be observed far from their gas nursery once they have passed by the extremities of the bar. This aspect makes the interpretation of observational data in this environment even more challenging than previously thought.

%%%%%
\subsubsection{Results}

\begin{figure}
\includegraphics{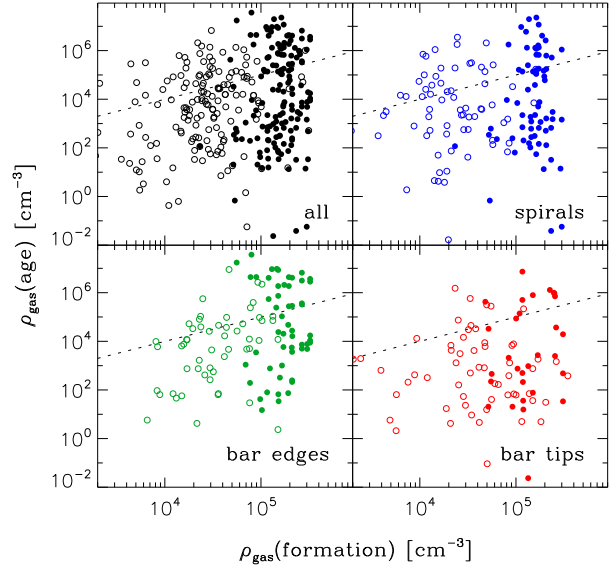}
\caption{Gas density at the position of the stars forming at $t = 790 \Myr$ (filled circles), and stars forming at $t = 770 \Myr$ (empty circles), taken at the time of their formation on the horizontal axis, and at $t = 800 \Myr$ on the vertical axis. Dotted lines indicate equality of the two densities. Only a fraction of stars is shown, for the sake of clarity. The top-left panel shows stars independently of their location in the galaxy, while the other three panels are subsets in specific formation regions (identified visually).}
\label{fig:densities_sn}
\end{figure}

\begin{figure}
\includegraphics{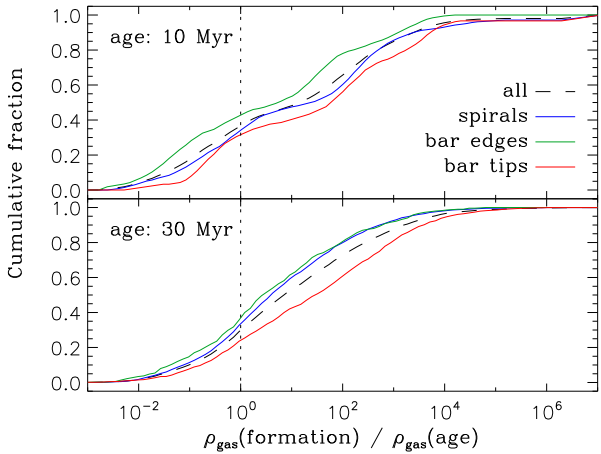}
\caption{Cumulative fraction of stars as a function of the ratio of ambient gas densities between the time of star formation and $10 \Myr$ (top) or $30 \Myr$ (bottom). In all cases, the majority of the stars experience a decrease of the ambient gas density, and this already after only $10 \Myr$.}
\label{fig:densities_sn_ratio}
\end{figure}

In \fig{densities_sn}, we compare the gas densities at the positions of stars (hereafter the ambient density) at the time of their formation with that $10$ and $30 \Myr$ later. In practice, we select stars younger than $1 \Myr$ at $t=770 \Myr$ and identify them at $t = 800 \Myr$. They constitute our $30 \Myr$-old sample. The ambient density is that of the gas cell in which the star particle is found. We repeat the exercise by selecting young stars at $t = 790 \Myr$ and identifying them at $t = 800\Myr$ to get our $10 \Myr$-old sample.

The evolution of the galaxy between $t = 770 \Myr$ and $t = 790 \Myr$ and the on-going collapse of clouds already noticed in \fig{ks} translate here into a rise of the average ambient density (i.e. an offset of the filled symbols to the right on \fig{densities_sn}). \fig{densities_sn_ratio} shows the cumulative fraction of the stars as a function of the ambient density ratio (density at formation divided by density $10$ or $30 \Myr$ later).

The gas density at the time of formation is fairly independent of the environment (spiral vs. bar). For about one third of the stars, the ambient density increases with time: either the star remains in the central region of its cloud which keeps collapsing, or the star orbit makes it fly through the central region. For the other two thirds, the ambient density drops as the stars age, as a result of feedback dissolving the cloud, and/or of the cloud-star decoupling. As discussed above, the decoupling effect is environment-dependent, and so is the ambient density drop.

The picture is made more complicated by the environmental dependence of the density around the cloud itself. Since the scale of molecular clouds is comparable to the tangential extent of a gaseous spiral arm, once a star leaves a cloud there, it also leaves the arm (see \citealt{Renaud2013b}, their Fig. 14). Therefore, its ambient density drops from that of the cloud centre ($\sim 10^{4\mh 6} \cc$ in our simulation) to that of the inter-arm region ($\sim 1\mh 10 \cc$) within a few $10 \Myr$. At the tip of the bar however, the inter-cloud medium remains relatively dense ($\sim 10^2 \cc$, recall \fig{density}). Therefore, when a star leaves its cloud there, it experiences a less abrupt change of its ambient density. As a result, the stronger decoupling effect in the bar than in spirals (\fig{decoupling}) is \emph{partially} balanced by a less abrupt variation of the ambient density. 

Stars formed at the edge of the bar experience an intermediate effect between those in the bar extremities and in the spirals. The rapid transit of stars along the bar edges due to the fast orbital motion on $x_1$-orbits also implies that they will experience the strong decoupling effect found at the tips within the first $\sim 20 \Myr$ of their life (recall \sect{velocity}).

In conclusion: compared to stars in spiral arms, a larger faction of the stars formed at the tip of the bar evolve into a low density medium (\fig{densities_sn_ratio}). Despite a denser inter-cloud medium, the ambient density after a given period of evolution ($10$ or $30 \Myr$) is on average lower in the bar than in spiral (\fig{densities_sn}). 

%%%%%
\subsubsection{Implications}

On top of the mere divergence in spatial distribution of the young stars relative to their birth places, the decoupling plays a potentially important role on the effects of feedback. In all regions, the decoupling effect is already measurable in the first $10 \Myr$ of the star life, i.e. even before the supernova (SN) blasts (set at $10 \Myr$ after star formation in our simulation). This implies that the feedback energy is not injected in the densest regions of clouds, but rather at their periphery or even in the inter-cloud medium. SN bubbles can then expand more easily in a relatively low density medium \citep{Chevalier1977}, and potentially drive galactic fountains or outflows. At first order, the net feedback effect on the ISM thus depends mainly on two factors: (i) how fast a star decouples from its cloud (\fig{decoupling}), and (ii) how dense the ISM surrounding the cloud is (\fig{densities_sn} and \ref{fig:densities_sn_ratio}).

This decoupling aspect is not retrieved in most of the other simulations. On the one hand, simulations of the ISM or of isolated clouds miss the large-scale dynamics of the galaxy and thus the very drivers of the decoupling (asymmetric drift, shear, tides). On the other hand, the decoupling scales ($\sim 10 \pc$) are not resolved by the simulations of large cosmological volumes. The injection of feedback is thus inaccurate in both cases.

The decoupling effect in bars that we noted could play a role in removing low angular momentum gas via outflows (or fountains) at early stages of galaxy formation, as proposed by \citet{Brook2012}. The low angular momentum gas, sitting at small galactic radii, ultimately gives rise to overly massive bulges and peaked rotation curves, in contrast to observed galaxies. This so-called ``angular momentum problem'' has plagued galaxy formation simulations for decades \citep[e.g.][]{Abadi2003, Navarro1994, Governato2007, Scannapieco2009, Ubler2014, Agertz2015}. Strong stellar feedback provides a solution, although this may have a detrimental effect on galactic thin disk formation \citep{Roskar2014}. By allowing for SN feedback to more efficiently blow out pockets of low density gas in specific galactic regions, cloud-star decoupling in bars could be an important ingredient in alleviating the problem.

Reality is much more complex than the picture we draw up here. The ISM is clumpy and non-uniform, making feedback bubbles asymmetric. Furthermore, clustered star formation implies that multiple, asynchronous, blast waves would interact in a given star forming volume. On top of this, colliding feedback shells and interaction with nearby clouds are likely to trigger star formation and thus subsequent injection of feedback in an already altered medium. A full understanding of the physics of feedback in galactic (or even cosmological) context requires more simulations like ours, which is still very costly. The zoom-in method by \citet{Bonnell2013} is likely to provide insights in this topic (see also \citealt{Rey-Raposo2015}).

%%%%%%%%%%%%%%%%%%%%%%%%%%%%%%%%%%%%%%%%%%%%%%%%%%%%%%%%%%%%%%%%%%%%%%%%%%%%%%%%
\section{Discussion and conclusions}

We use a hydrodynamical simulation of a Milky Way-like galaxy to probe the role of the bar environment in triggering and regulating the formation of clouds and stars. Our approach consists in connecting kpc-scale processes with the formation and early evolution of parsec-scale structures. Our main conclusions are:
\begin{itemize}
\item The leading edges of the bar yield converging flows and large-scale shocks that favour the gathering of dense gas in kpc-long filaments, confirming the results of \citet{Athanassoula1992} and others. Furthermore, these regions host supersonic turbulence which then helps the fragmentation of these filaments into clouds.
\item The tangential velocity gradient at the edge of the bar is much weaker than in innermost regions, while the radial component yields shocks. Such differences explain that gas clumps are likely to survive at the edge but not inside the bar.
\item The orbital pattern in the bar creates a fast circulation of matter along the bar. While going from one tip to the other takes only $20 \Myr$, the gas slows down at both tips and accumulates there.
\item Because of the high velocities along the bar, the objects older than $\sim 10 \Myr$ have formed in a different environment and under different physical conditions than the ones they are when observed.
\item Orbital crowding at the tip of the bar leads to cloud-cloud collisions that can form massive molecular complexes like W43 and massive stellar associations ($\sim 10^6 \Msun$). Such objects remain at the tip of the bar for a rather short period ($\sim 10 \Myr$ or less) after their formation.
\item A cloud-cloud collision increases the rotation of the merger, the turbulent Mach number inside the cloud and the efficiency of star formation, in a comparable way as galaxy mergers.
\item Young stars decouple from their gas cloud more efficiently at the tips of the bar than along the edge or in spiral arms. Such aspect can affect the net effect of stellar feedback, possibly even at galactic scale.
\end{itemize}

All together, these points illustrate the paramount effect of the fast dynamical evolution connected to the bar: stars even only a few $10 \Myr$ old can be found in a very different environment than the one they formed in. Tips of the bar play an important part in this picture. Either they host clouds long enough to witness their collapse and star formation, or they gather physical conditions allowing star formation events (like cloud-cloud collisions). In the latter case, the actual event can occur a few Myr later, while the clouds already exited this particular region.

The processes we highlight are likely to be altered in weaker bars \citep[but see][and references therein]{Zhou2015} or by the presence of multiple bars, especially if they rotate at different speed \citep{Wozniak1995}, e.g. by changing the large scale velocity field. Active star forming regions, comparable to those we analysed, has been observed at the extremities of bars of several galaxies, like NGC~1097 \citep{Ondrechen1983}, NGC~1433, NGC~1512 and NGC~5383 \citep{Athanassoula1992}.

\citet{Kraljic2012} showed that, although bars are short-lived at high redshift ($z >1$), those formed at $z\approx 0.8 \mh 1$ survive for long periods. They evolve through the accretion of intergalactic gas that they re-distribute within the galactic disc via gravitational torques \citep{Bournaud2005}. According to the classification of bars in isolated galaxies proposed by \citet{Verley2007}, the case modelled here is representative of the most common class (their group E). The evolution scenario in \citet{Verley2007} suggests such bar would weaken and even be destroyed by gas infall. Therefore, simulations at parsec resolution and in cosmological context are needed to establish the frequency and the importance of the aspects we presented here in the special case of our Milky Way-like simulation.

%%%%%%%%%%%%%%%%%%%%%%%%%%%%%%%%%%%%%%%%%%%%%%%%%%
\section*{Acknowledgments}
We thank the reviewer for a report that helped improve the clarity of the paper. This work was granted access to the PRACE Research Infrastructure resource \emph{Curie} hosted at the TGCC, France (PRACE project ra-0283), and national GENCI resources (projects 2013, 2014, 2015-GEN2192). FR and FB acknowledges support from the European Research Council through grants ERC-StG-335936 and ERC-StG-257720. EA acknowledges financial support from the People Programme (Marie Curie Actions) of the European Union's Seventh Framework Programme FP7/2007-2013/ under REA grant agreement number PITN-GA-2011-289313 to the DAGAL network and from the CNES (Centre National d'Etudes Spatiales - France). FC acknowledges the European Research Council for the Advanced Grant Program Num 267399-Momentum. KK acknowledges support from grant Spin(e) ANR-13-BS05-0005 of the french ANR.

%%%%%%%%%%%%%%%%%%%%%%%%%%%%%%%%%%%%%%%%%%%%%%%%%%
\appendix
\section{Computation of the decoupling force}
\label{app:decoupling}

To measure the effect of the decoupling of the stars from the gas cloud in which they form, we first identify clouds as density peaks in one snapshot of the simulation. We select the grid cells in which the gas is denser than the star formation threshold ($2000 \cc$). All contiguous selected cells constitute one cloud. Then, we select stars between 2 and $10 \Myr$ old, and identify their closest cloud as their formation site. Choosing older stars could lead to errors in the identification of the formation site. We compute the velocity of the star with respect to the cloud and its radial and tangential component. We reject the stars with a negative radial velocity, as they are moving toward the cloud and do not experience a decoupling effect. The total velocity is then converted into a force, by multiplying it by the mass of the star, and dividing it by the age of the star. Finally, we compute the average force for each cloud over all its stars.

To obtain the map of \fig{decoupling}, we convolve the scattered distribution of the force for each cloud by a Gaussian kernel. We then multiply this map by a smoothed version of the gas density map for graphical purposes only.

%%%%%%%%%%%%%%%%%%%%%%%%%%%%%%%%%%%%%%%%%%%%%%%%%%

\bibliographystyle{mn2e}

\end{document}